\documentclass[aps,prl,nofootinbib,twocolumn,floatfix,superscriptaddress]{revtex4}
\usepackage{latexsym} \usepackage{amsmath} \usepackage{dcolumn}
\usepackage{bm} \usepackage{graphicx} \usepackage{epsfig}
\usepackage{float} \usepackage{stackrel,amssymb}
\usepackage{slashbox}
\usepackage{float}
\usepackage{url}
\graphicspath{{./Figures/}}

\usepackage[]{graphicx}
\usepackage[]{wrapfig}
\usepackage[]{color}
\begin{document}

\title{
Vibrational Signatures in the THz Spectrum of $1,3$-DNB:
A First-Principles and Experimental Study}

\author{Towfiq Ahmed}
\email{atowfiq@lanl.gov}
\affiliation
{Theoretical Division,
Los Alamos National Laboratory, Los
Alamos, New Mexico 87545}
\author{Abul K. Azad}
\email{aazad@lanl.gov}
\affiliation
{Center for Integrated Nanotechnologies,
Los Alamos National Laboratory, Los
Alamos, New Mexico 87545}
\author{Raja Chellappa}
\affiliation
{Lujan Neutron Scattering Center,
Los Alamos National Laboratory, Los
Alamos, New Mexico 87545}
\author{Amanda Higginbotham-Duque}
\affiliation
{Shock and Detonation Physics,
Los Alamos National Laboratory, Los
Alamos, New Mexico 87545}
\author{Dana M. Dattelbaum}
\affiliation
{Shock and Detonation Physics,
Los Alamos National Laboratory, Los
Alamos, New Mexico 87545}
\author{Jian-Xin Zhu}
\affiliation
{Theoretical Division,
Los Alamos National Laboratory, Los
Alamos, New Mexico 87545}
\affiliation
{Center for Integrated Nanotechnologies, Los Alamos
National Laboratory, Los Alamos, New Mexico 87545}
\author{David Moore}
\affiliation
{Shock and Detonation Physics,
Los Alamos National Laboratory, Los
Alamos, New Mexico 87545}
\author{Matthias J. Graf}
\affiliation
{Theoretical Division,
Los Alamos National Laboratory, Los
Alamos, New Mexico 87545}
\affiliation
{Office of Science, U. S. Department of Energy, Washington, DC 20585-1290}

\date{\today}
%%%%%%%%%%%%%%%%%%%%%%%%%%%%%%%%%%%%%%%%%%%%%%%%%%%%%%%%%%%%%%%%%%%%%
%% The document title should be given as usual
%% A short title can be given as a suggestion for running headers.
%%%%%%%%%%%%%%%%%%%%%%%%%%%%%%%%%%%%%%%%%%%%%%%%%%%%%%%%%%%%%%%%%%%%%

%%%%%%%%%%%%%%%%%%%%%%%%%%%%%%%%%%%%%%%%%%%%%%%%%%%%%%%%%%%%%

\begin{abstract}
Understanding the fundamental processes  of light-matter interaction is important for detection of explosives and other energetic materials, which are active in the infrared and terahertz (THz) region. We report a comprehensive study on electronic and vibrational lattice properties of structurally similar $1,3$-dinitrobenzene ($1,3$-DNB) crystals through first-principles electronic structure calculations and THz spectroscopy measurements on polycrystalline samples. Starting from reported x-ray crystal structures, we use density-functional theory (DFT) with periodic boundary conditions to optimize the structures and perform linear response calculations of the vibrational properties at zero phonon momentum. The theoretically identified normal modes agree qualitatively with those obtained experimentally in a frequency range up to 2.5 THz and quantitatively at much higher frequencies. The latter frequencies are set by intra-molecular forces. Our results suggest that van der Waals dispersion forces need to be included to improve the agreement between theory and experiment in the THz region, which is dominated by intermolecular modes and sensitive to details in the DFT calculation.  An improved comparison is needed to assess and distinguish between intra- and intermolecular vibrational modes characteristic of energetic materials.

%We performed first-principles electronic structure and lattice vibrational calculations of 1,3-dinitrobenzene (1,3-DNB) crystals
%to compare with terahertz measurements with the goal to  identify the characteristic absorption signature of electromagnetic waves up to 3 THz.
%We calculated the volume dependence of the total energy of crystalline 1,3-DNB using the density functional theory, from which the corresponding vibrational modes were calculated at the measured and theoretically relaxed structures.  A comparison with measured modes showed qualitative agreement within 0.5 THz at low frequencies.
%The calculated modes of the molecular crystal were further validated and analyzed through a comparison with first-principles calculations of a single 1,3-DNB molecule in a large cell with periodic boundaries.
%%and finally with an isolated single molecule.
%This comparison allowed a distinction between intra- and intermolecular vibrational modes as well as the sensitivity of these  modes to different theoretical approximations.
%Since 1,3-DNB is physically and structurally similar to many explosives and energetic materials, identifying materials specific signature in the THz region through first-principles methods is important for understanding the fundamental processes of light-matter interaction for their detection.
\end{abstract}

%\pacs{78.70.Dm, 71.10.Fd, 71.10.-w, 71.15.Qe}

%\keywords{THz, DFT, LDA, vibrations, explosives}
\maketitle
%%%%%%%%%%%%%%%%%%%%%%%%%%%%%%%%%%%%%%%%%%%%%%%%%%%%%%%%%%%%%

\textit{Introduction.~} The challenging problem of reliable detection of energetic materials and explosives is an evolving and active research field. A particularly interesting
aspect is the remote stand-off detection using the molecular fingerprint in the THz region~\cite{chen_2004,jian_2007,Lee_book}.
On the one hand, THz time-domain spectroscopy (THz-TDS) has shown in the past decade  to be a powerful method to characterize the vibrational modes in the 0.1-3.0 THz region and thus provides the needed spectral fingerprints. On the other hand, predicting and identifying the vibrational modes of energetic materials from first-principles density functional theory (DFT) calculations in the THz regime remains a daunting challenge~\cite{shimojo_2010}. Small differences in the energy of the electronic structure calculations on the order of 1 meV can have drastic effects on the low frequency vibrational spectrum in the terahertz region (1 THz $\approx$ 4.14 meV). This is especially true for energetic materials and explosives, which have a quite complicated molecular crystal structure.
Therefore, a success in overcoming this challenge will have huge technological implications. In particular, knowing the vibrational modes of a dielectric insulator will allow us to understand the origin of the electromagnetic signature of the molecular crystal for fingerprinting its response in the THz regime of these materials.

The $1,3$-DNB molecular crystal is an inert material, yet an ideal substitute for explosives due to its similarity in chemical composition, molecular structure, and physical and dielectric properties. It also shares many common structural properties with explosives such as a backbone benzene ring, NO$_2$ antenna groups, intermolecular hydrogen bonds, and planar packing.  Here we investigate the vibrational spectrum of the $1,3$-DNB molecular crystal using periodic DFT and THz-TDS methods.  Our DFT calculations on a model $1,3$-DNB single molecule with periodic boundary conditions identify most of the normal modes, which were measured accurately by our THz-TDS technique on the more complicated molecular crystal within the frequency between 0.5 and 2.5 THz. Interestingly, the
prediction  also reproduces, for the first time, the lowest energy mode at $\approx$ 0.45 THz, which is in good agreement with our
experiments and earlier work by Fischer {\it et al.}~\cite{fischer_2007}.
Our work demonstrates that DFT can provide a theoretical underpinning for an accurate detection of
$1,3$-DNB and similar  explosive  materials.
The experimental validation of DFT predictions should drastically improve once van der Waals dispersion forces are included to deal with intermolecular interactions.
Finally, this suggests that such
theoretical capability should also enable
the identification of intrinsic THz modes from high-order harmonics of GHz modes, which can originate from intermolecular interactions.

\begin{figure}[!ht]
\includegraphics[width=3.5in]{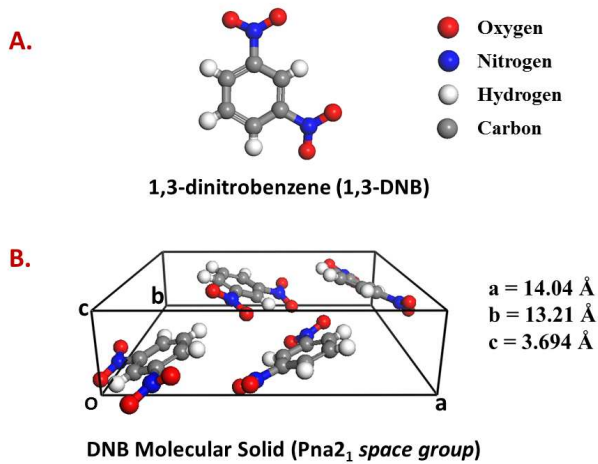}
\caption{{\bf {a}} Schematic image of the
1,3-DNB
single molecule.
{\bf {b}} Reported x-ray crystal structure of the
1,3-DNB molecular solid with
experimental lattice parameters at 130 K after Ref.~\cite{wojcik_2002}.
}
\label{f1}
\end{figure}

\textit{Methods.~}  All calculations were performed by using the plane-wave pseudo-potential method of DFT with periodic boundary conditions, as implemented in VASP~\cite{vasp} and Quantum Espresso (QE)~\cite{espresso} codes.  The crystal structure of  the $1,3$-DNB solid, as shown in Fig.~\ref{f1}(b),  is orthorhombic  (space group Pna2$_1$ or \#33)~\cite{wojcik_2002,cod1,cod2,cod3,url1}.  Each unit cell contains four $1,3$-DNB molecules and at 130 K has the reported linear dimension of
$a=14.040$ \AA, $b=13.208$ \AA, $c=3.6940$ \AA~\cite{wojcik_2002}.
% {\color{red} [Towfiq, please clarify the value difference here from
% that listed in Fig.~\ref{f1}(b).]}
We used the x-ray refined atomic structure provided in
the Crystallography Open Database~\cite{cod1,cod2,cod3,url1} as the starting basis for our studies.

We considered vibrational properties for both an isolated $1,3$-DNB single molecule by introducing an enlarged periodic supercell of dimension ($a\times b \times 4c$) and the molecular crystal case. For the single-molecule case, as shown in Fig.~\ref{f1}(a),  the calculations were performed by using the QE 5.0.1 package with the provided  ultra-soft
PBE~\cite{pbe} pseudo-potentials~\cite{pseudo1}. The first Brillouin zone is sampled with $8 \times 8 \times 8$ $k$-points. The energy cutoffs of
40 and 160 Rydberg (Ry) were taken for wave function and charge density, respectively. With the fixed linear dimension of the supercell, the internal atomic structure was relaxed and optimized
with the self-consistency resolution of $10^{-14}$ Ry
%{\color{red} [
%**Ry** Towfiq, please provide the unit of energy resolution.]}
for total energies and 0.4 mRy/$\AA$ for mean atomic forces per atom.

For the $1,3$-DNB molecular crystal, as shown in Fig.~\ref{f1}(b),  we used the VASP-5.3.1 package~\cite{vasp} to calculate the total energy as a function of volume and optimized the internal atomic positions at each volume.  For all atoms,  the ultra-soft~\cite{pseudo1,pseudo2} PBE pseudo-potentials, provided by  VASP, is used and the first Brillouin zone is sampled with a mesh of $6 \times 6 \times 6$ $k$-points.  The energy and force convergence criteria are $10^{-6}$  eV and $10^{-4}$ eV/$\AA$ per atom, respectively.
%{\color{red} [**eV** Towfiq, please check the values and their units here.]}
Various values of the energy cutoff parameter, 500 eV and 700 eV, were taken for the reliability of the optimized volume.  For both single molecule and
molecular
crystal case, the density functional perturbation theory (DFPT)
was used to obtain in linear response
the vibrational dispersion at the center of the lattice Brillouin zone
(phonon momentum $\mathbf{q}=0$. For the latter case we also used
the Phonopy package~\cite{phonopy} for crosschecking the results, and confirmed that the calculated mode frequencies obtained from both approaches
agreed well with each other.

\begin{figure}[t]
\centering\includegraphics[width=1.0\linewidth,clip]{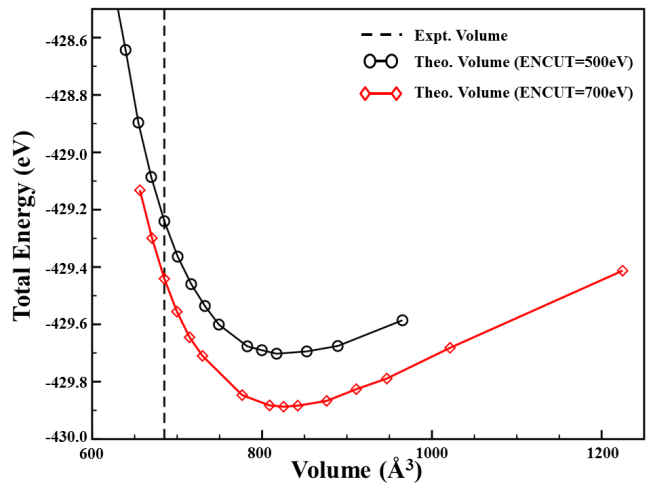}
\caption{
Comparison between experimental and theoretical volume of 1,3-DNB molecular crystalline
solid. The energy-volume curves,
obtained using DFT with PBE functional at absolute zero temperature,
are shown by the solid black and red curves for two energy cutoff values of
500 eV and 700 eV, correspondingly. The theoretically found equilibrium position
is around 825 \AA$^3$, and is significantly larger than the experimentally observed
value of 685 \AA$^3$ at 130 K,
marked by a vertical dashed line.
}
\label{f2a}
\end{figure}

The sample was prepared from commercially available $1,3$-DNB powder and hand-pressed at 4000 psi using a 2-ton Carver press to 6 mm diameter by 1 mm thick pellets.
A metallic sample holder containing two
identical apertures of diameter 3.5 mm is used for sample measurements.\cite{azad1} During
measurements, the sample is attached to one hole, while the other clear hole
is used as the reference. We employed a ZnTe-based THz-TDS spectrometer for sample measurements.\cite{azad2}
A Titanium:Sapphire amplified pulsed laser system, operating at $\lambda$=800
nm with 50 fs laser pulses, is used for THz generation and detection via nonlinear electro-optic effect in ZnTe
crystals. Generated THz pulses are collected, collimated through the sample,
and finally detected on the THz detector by two pairs of parabolic mirrors.
The THz beam centered through the sample holder's aperture has a frequency
independent diameter of 3.5 mm. To mitigate the THz absorption by the ambient
water vapor, the enclosed THz system is continuously purged with dry air.
The sample holder is attached to a liquid helium cooled cryostat, allowing
precise control of sample temperature. The measured time-dependent THz pulses
transmitting through the
sample and reference are converted to complex frequency
spectra using numerical Fast Fourier Transform. Because of the finite thickness of
the
sample pellets,
the THz pulse through the sample is usually
accompanied by multiple reflected pulses. However, clear time separation
between main and reflected pulses enables truncation of the sample THz
pulses just before the arrival of the first reflection, alleviating
multi-reflection-induced oscillations in spectra, but also limiting the low frequency resolution.

\textit{Results and discussions.~}  Our calculations for the $1,3$-DNB single molecule with periodic boundaries
gives all normal modes with physically real frequencies, except for the three translational modes. For the $1,3$-DNB  molecular crystal,
the calculations yield imaginary vibrational modes, irrespective of using the reported x-ray crystal structure or
the relaxed atomic structure obtained with fixed experimental lattice parameters~\cite{wojcik_2002} at 130 K,
suggesting the structure is dynamically unstable toward small perturbations.
Therefore, we performed total energy calculations for a sequence of volumes. For each of them the internal atomic positions are fully optimized.   Figure~\ref{f2a} shows the volume dependence of the total energy calculated with an energy cutoff of 500 eV and 700 eV. A rigid shift in energy can collapse both curves, demonstrating good convergence for finding the optimized volume. The theoretically optimized  unit cell has the linear dimension of $a_{\text{opt}}=s a$, $b_{\text{opt}}=s b$, and $c_{\text{opt}}=s c$, where uniform relaxation of the lattice parameters with scaling factor $s=1.064$ was assumed. This scaling amounts to a theoretical unit volume of 825 \AA$^3$, which is
noticeably bigger than that of the
experimental
 volume of 685 \AA$^3$~\cite{wojcik_2002} (marked by a vertical dashed line in Fig.~\ref{f2a}).
 The overestimation of the volume could be due to the PBE exchange-correlational functional, which routinely
underestimates the binding strength of molecular crystals. Similar discrepancies have
been reported in the literature for other molecular crystals~\cite{Montanari1998, Todorova2010}. This behavior is indicative of the importance of van der Waals dispersion forces, which are absent in our DFT calculation.
Noteworthy, we found that the calculations with the theoretically optimized volume give all normal vibrational modes.

 \begin{figure}[t]
\centering\includegraphics[width=1.0\linewidth,clip]{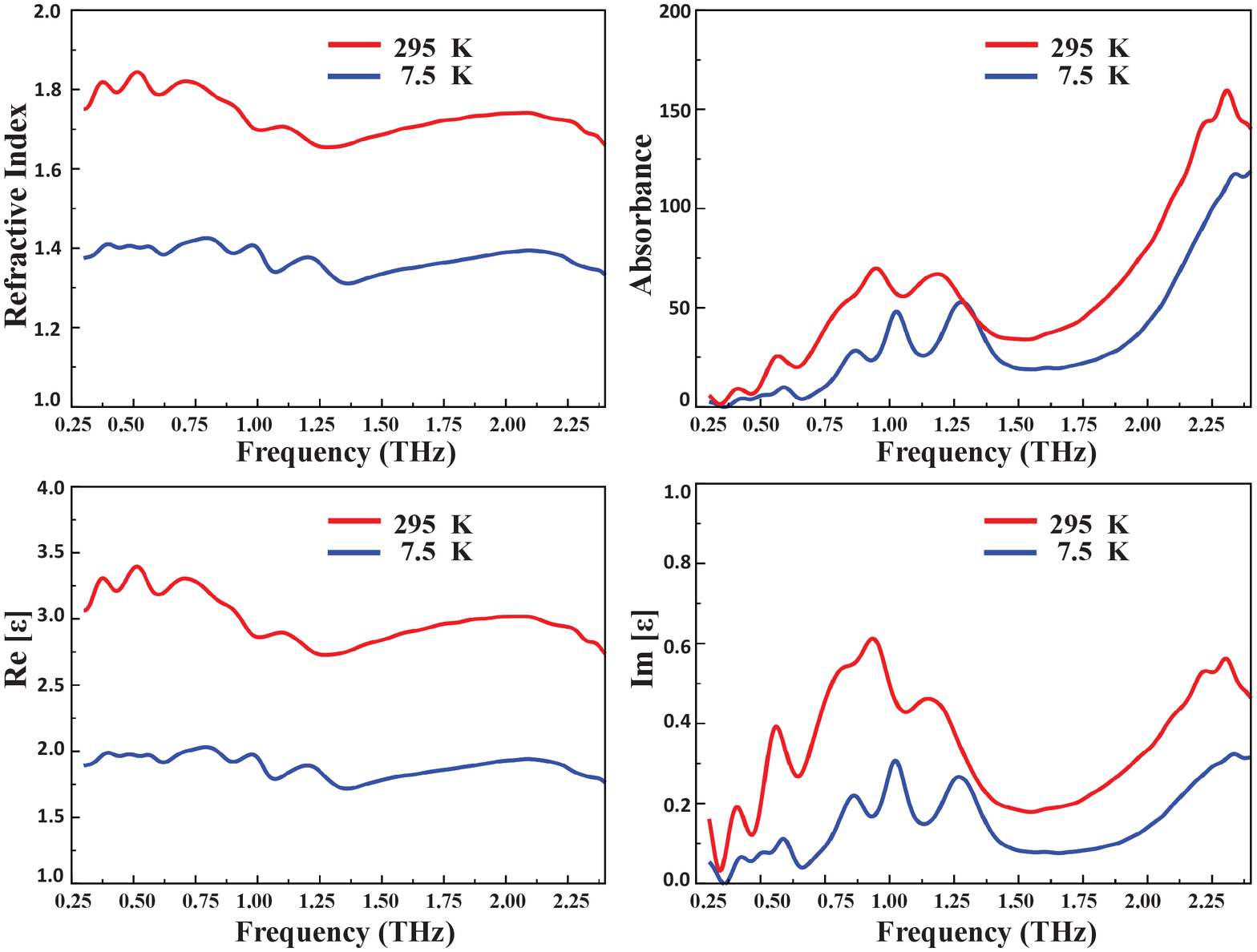}
   %\begin{minipage}[!t]{0.50\linewidth}
   %\epsfig{file=Fig3.eps, width=\linewidth}
   %\end{minipage}\hfill
   %\begin{minipage}[!t]{0.48\linewidth}
\caption{
Frequency dependence of the
 refractive index {\bf {(a)}},
 absorbance {\bf {(b)}},
real {\bf {(c)}} and imaginary {\bf {(d)}}  parts of the dielectric function, extracted from the THz-TDS experiments measured on a polycrystalline
(compressed powder) pellet
$1,3$-DNB at 295 K (red) and 7.5 K (blue) curves, respectively.
}
\label{f3}
%\end{minipage}
\end{figure}

After Fast-Fourier transforming our measured THz-TDS spectra, we
fit the complex spectra to the following relation between the electric field of the sample and reference at normal incidence~\cite{azad2}
\begin{equation}\label{Fresnel}
\frac{E_{smpl}(\omega)}{E_{ref}(\omega)}=t_{12}t_{21}e^{i(k-k_0)d}e^{-\alpha d/2}\;,
\end{equation}
to obtain the complex refractive index $n= n_r + i n_i$ and the absorption coefficient  $\alpha = 4\pi n_i/\lambda$.
Here  $k_0 = 2\pi/\lambda $ is the free space wave vector, $\omega=k_0 c$ with speed of light $c$,
$k=2\pi n_r/\lambda$ is the wave vector inside the sample,
$d$ is the sample thickness, and t$_{12}(\omega)$ and t$_{21}(\omega)$ are the
frequency-dependent Fresnel transmission coefficients from the front- and
back-side of the sample.  With the extracted complex refractive index and absorption coefficient,  the complex dielectric function is given through the relation
\begin{eqnarray}
\epsilon(\omega) &=& \epsilon_r + i\epsilon_i = (n_r + i n_i)^2
\nonumber\\
&=& (n_r + i\alpha \lambda/4\pi)^2\;.
\end{eqnarray}
That is,    $\epsilon_r=n_r^2 - (\alpha \lambda/4\pi)^2$ and $\epsilon_i=\alpha \lambda/2\pi$.

%Generally, the complex dielectric
%function $\epsilon(\omega)$ of materials consists of dielectric contributions
%from the conduction electrons, lattice vibrations (optical phonons),
%and high-frequency dielectric constant $\epsilon_{\infty}$.

Figure~\ref{f3} shows the extracted refractive index, absorbance (absorption coefficient), and the complex dielectric function (real and imaginary parts)
from the measured THz spectra on $1,3$-DNB polycrystal at
temperatures 295 K and 7.5 K.
%At low temperatures the refractive index is reduced by roughly 20\%
%due to lattice anharmonicity and all absorption peaks are blue shifted to higher frequencies,
%caused by the contraction of the solid.
In our experiment, the absorption spectrum shows multiple signatures of prominent resonant peaks within our
measured spectral window between 0.3 and 2.5 THz.
As shown in Fig.~\ref{f3}(a) and (b), both refractive index and absorbance show weak frequency dependence, yet strong
temperature dependence. Their values decrease as the temperature of the
sample decreases
from room temperature to 7.5 K.
The refractive index drops nearly
30\%, while the real part of the dielectric function drops roughly 60\%.
In addition, we observe
blue shifts and narrowing
of the absorption peaks at cryogenic temperature.

\begin{figure}[t]
\centering\includegraphics[width=1.0\linewidth,clip]{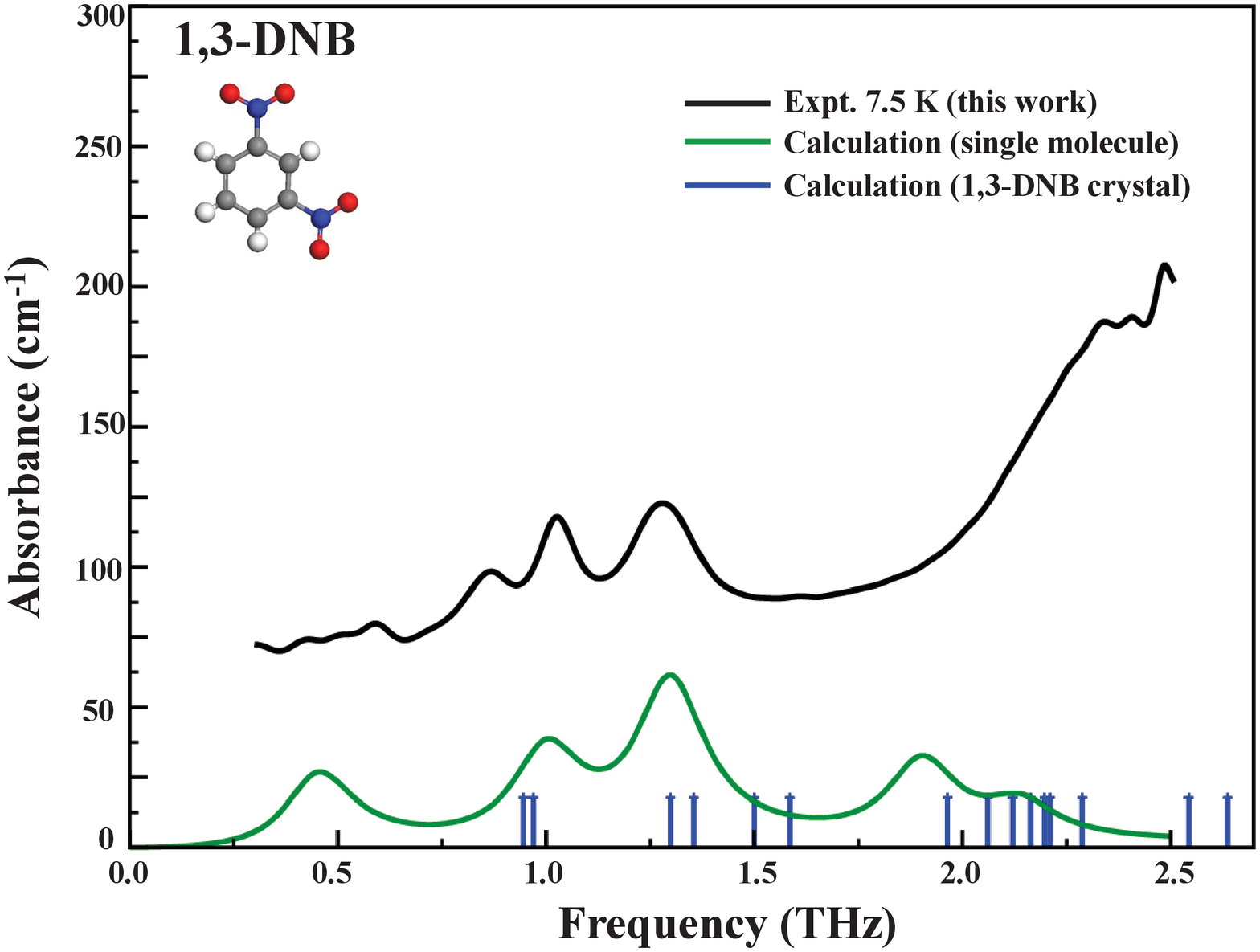}
 %  \begin{minipage}[!t]{0.50\linewidth}
 %   \epsfig{file=Fig4.eps, width=\linewidth}
 %  \end{minipage}\hfill
 %  \begin{minipage}[!t]{0.48\linewidth}
\caption{
Comparison between experimental and theoretical THz absorbance spectra;
measurements (solid black lines) were performed
at 7.5 K. Calculations are shown for both single molecule (solid green),
and crystalline solid (vertical blue) of 1,3-DNB.
For the crystalline solid,
only the phonon frequency modes are calculated, which are identified
by the vertical lines.
%for comparison, we also present earlier experimental
%results  (dashed magenta line) by Fischer {\it {et al.}}\cite{fischer_2007} and
%(dashed red line) by Lee {\it {et al.}}\cite{Lee_book}
}
\label{f4}
%\end{minipage}
\end{figure}

In Fig.~\ref{f4}
the first-principles
phonon modes for the molecular crystal
are shown as vertical solid lines.
Comparing our crystalline $1,3$-DNB calculation with our THz experiment
(solid black curve), we find the observed
modes at 1.03 THz and 1.28 THz  are qualitatively identified,
while the 0.59 THz normal mode
is missing from our calculations. On the other hand, in our DFPT
phonon calculation
for the periodic
single molecule of 1,3-DNB (solid green curve, obtained via Eq.~(\ref{Epsilon}) below), using the PHonon package of QE \cite{espresso}
we recover the lowest
frequency mode at 0.45 THz. Our calculated lowest frequency has a
prominent and stronger
oscillator strength compared to our experimental observation
at 0.59 THz, while it is in better agreement with the absorbance peak reported by Fischer~\cite{fischer_2007}. The origin of this discrepancy is unclear and difficult to sort out, since Fischer mixed polyethylene with 1,3-DNB to press pellets of unspecified thickness.
Our interpretation is
that such modes  possibly originate from the librational
motion of the NO$_2$ antenna groups
attached to the `1' and `3' positions of the carbon ring. A different conclusion was drawn from a rigid structure model analysis by W{\'{o}}jcik {\it {et al.}},
who assigned these internal torsion modes to 1.89 and 1.92 THz.\cite{wojcik_2002}
In the crystalline phase with a more ordered and rigid configuration, such motions are constrained and would explain the
weaker oscillator strength in the infrared
absorbance spectra measured in
this work.

\begin{table*}
\caption{ Experimental and DFPT calculated THz frequency modes and oscillator
strengths of 1,3-DNB molecular crystal. The oscillator strengths for the single molecule (sm) obtained from QE
were scaled by factor four to account for the four molecules per unit cell, and uniform damping of 0.2 THz was assumed. }
\centering
\begin{tabular}{|c||c c c||c c||c||c| }
\hline
Mode  & $\omega_j/2\pi$ (THz) & $f_j$ & $\Gamma/2\pi$ (THz) & $\omega_j/2\pi$ (THz) & $f_j$ & $\omega_j/2\pi$ (THz) &  $\omega_j/2\pi$ (THz)\\[0.5ex]
&  (Exp) & (Exp) & (Exp)	& (DFT sm) & (DFT sm) & (DFT x-tal) & (W\'ojcik)\\[0.5ex]
\hline
1 & 0.59 & 0.012 & 0.07	& 0.45 & 0.175 &       	& 0.57	\\[0.5ex]
2 & 0.86 & 0.031 & 0.15	&  	& 		& 0.95		&  0.72, 0.84	\\[0.5ex]
3 & 1.03 & 0.026 & 0.11	& 1.00 & 0.044 & 0.97	&  0.93, 0.96	\\[0.5ex]
4 & 1.28 & 0.037 & 0.20	& 1.30 & 0.043 & 1.29	&  1.14	\\[0.5ex]
  &      &       &			&      &       	& 1.35		&  1.20  	\\[0.5ex]
  &      &       &			&      &       	& 1.50		&   1.68  \\[0.5ex]
  &      &       &			&      &       	& 1.59		&     	\\[0.5ex]
5 & 2.40 & 0.090 & 0.60	& 1.90 & 0.010 & 1.96	 & 1.83	\\[0.5ex]
  &      &       &			& 2.14 & 0.003 & 2.06	&   2.34  \\[0.5ex]
  &      &       &			&      &       	& 2.12		&   2.37  \\[0.5ex]
  &      &       &			&      &       	& 2.16		&     	\\[0.5ex]
  &      &       &			&      &       	& 2.20		&     	\\[0.5ex]
  &      &       &			&      &       	& 2.29		&     	\\[0.5ex]
  &      &       &			&      &       	& 2.54		&     	\\[0.5ex]
\hline
\end{tabular}
\label{table}
\end{table*}

Furthermore,  in the absence
of a contribution from conduction electrons, which is the case for the dielectric
$1,3$-DNB molecular solid, the total complex dielectric function can be
described by the Lorentz model~\cite{azad3}
\begin{equation}
\label{Epsilon}
\epsilon(\omega)=\epsilon_{\infty} + \sum_j \frac{f_j \omega_j^2}{\omega_j^2-\omega^2-i \Gamma_j \omega} \;.
\end{equation}
Here $\epsilon_{\infty}$ is the high-frequency dielectric constant,  $f_j$ is the dimensionless oscillator strength, $\omega_j$ is the $j$-th normal mode, and
$\Gamma_j$ is the damping constant. For simplicity, we assume frequency independent damping.
By fitting  the measured $\epsilon(\omega)$ at 7.5 K to Eq.~(\ref{Epsilon}),
we identify the first five transverse optical (TO) phonons, including their oscillator strengths and
damping constants, which are listed together with the different
theoretical values of low-energy modes in Table~1.

\begin{figure}[t]
\centering\includegraphics[width=1.0\linewidth,clip]{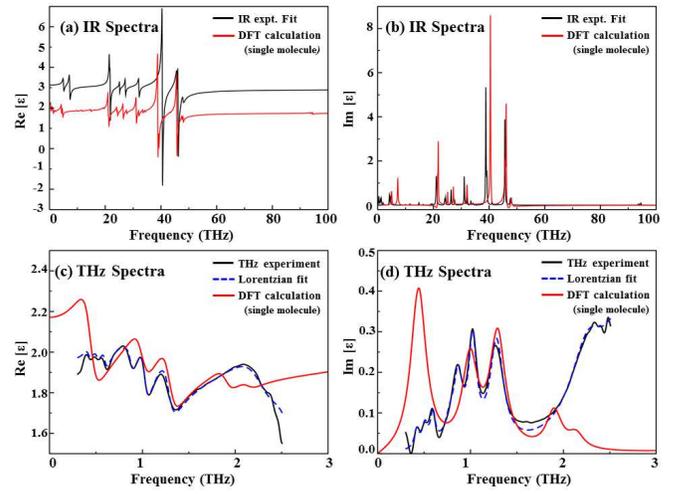}
  % \begin{minipage}[!t]{0.70\linewidth}
  %  \epsfig{file=Fig5.eps, width=\linewidth}
  % \end{minipage}\hfill
  % \begin{minipage}[!t]{0.28\linewidth}
\caption{
Comparison between interpolated experimental infrared (IR)
 spectra at 130 K \cite{TG_2010} (solid black) and theoretical (solid red) single molecule calculation for 1,3-DNB.
The theoretical complex dielectric function used the single
fit parameter $\epsilon_\infty=1.75$ to fit the THz spectrum at 7.5 K.
Panels
{\bf{(a)}} and {\bf{(b)}} show
the Re[$\epsilon(\omega)$] and Im[$\epsilon(\omega)$] in the
infrared region up to 100 THz in the
far-infrared (high frequency) region;
{\bf{(c)}} and {\bf{(d)}} show the low-frequency THz spectra
of
Re[$\epsilon(\omega)$] and
Im[$\epsilon(\omega)$] for
 the calculations (solid red), THz-TDS experiments (solid
black), and Lorentz fit to experimental data (dashed blue).
}
\label{f5}
%\end{minipage}
\end{figure}

By setting the damping coefficient to a constant for all modes (a reasonable physical value is 0.2 THz), we can calculate the dielectric function and in turn the
absorption coefficient for the single molecule (e.g., the green curve in Fig.~\ref{f4}).
We find that the calculated real and imaginary dielectric functions agree reasonably well with our
measurements at low temperatures with only a single fit parameter $\epsilon_\infty$. The direct
comparison is shown in Fig.~\ref{f5}.
For a consistency check we also compare our calculated real and imaginary dielectric functions for the single molecule with those of
single-crystal experiments reported by Trzebiatowska-Gusowska {\it {et al.}}\cite{TG_2010}
Our calculated complex dielectric function in the near- and far-infrared regions is in good agreement
 with the interpolation
function reported for the ten most prominent absorption signatures in Ref.~\cite{TG_2010},
as shown in
~\ref{f5} (a) and (b).
The largest mode in the single-molecule DFPT calculation is 94.5 THz and compares favorably with the measured IR mode at 93.3 THz.
We also successfully reproduce the large measured
spectral gap \cite{TG_2010}
between 51 and 86 THz
(vs.~50 and 93 THz) without adjustable parameters.
The offset seen in Re~$\epsilon$
between the experimental IR curve and theory stems from different parameters $\epsilon_\infty$ used at room
temperature and  7.5 K, due to
its strong temperature dependence.
Overall the high frequency intra-molecular modes of the single molecule are well captured by the DFPT method, despite the lack of van der Waals dispersion forces in the periodic DFT calculations and temperature differences between theory and experiment.

In the THz region, where intermolecular interactions are expected to play a dominant role, our single-molecule
calculations show qualitative agreement with both the
real, \ref{f5}(c), and imaginary, \ref{f5}(d), dielectric functions obtained from our pressed powder pellet.
The magnitude of the distinct peak-like feature present in our calculation at 0.45 THz is
stronger than the experimental peak intensity at 0.59 THz. This peak suppression may be
due to the surrounding environment of inter-molecular forces in the crystal, which constrains the motion of the NO$_2$ antenna group present
in the individual molecules of the solid. Our measured peak
at 0.86 THz was not captured in the DFPT calculation of the single molecule, while it agrees with the lowest mode at 0.95 THz in the crystal. It is feasible that incorporation of van der Waals interactions lowers the symmetry of the crystal and generates additional
low-frequency modes missed currently.

The five lowest frequency vibrational modes, which are observed in our measurements and calculations, are presented in Table~1.
They compare reasonably well with the infrared and Raman measurements
at 0.57, 0.72, 0.84, 0.93, 0.96, 1.14, 1.20, 1.68, 1.83, 2.34, 2.37, 4.89 and 5.94 THz
reported originally by Bobrov {\it et al.}\cite{Bobrov_1974} and revisited by W{\'{o}}jcik {\it {et al.}}\cite{wojcik_2002}, who used
a rigid structure analysis for
the normal modes of the single molecule.

\textit{Conclusions.~} In summary, we performed first-principles DFT electronic structure and DFPT phonon calculations of the THz spectrum of the dielectric 1,3-DNB molecular crystal
to compare with THz-TDS measurements in the 0.3-2.5 THz region.
Calculations for the single molecule with periodic boundaries and the molecular crystal were in overall good agreement except for the low-frequency THz modes of interest, pointing to subtle differences in intermolecular interactions and details in DFT codes.
The relaxed structural crystal properties obtained from VASP differ
noticeably from the experimentally reported volume, leading to a larger unit cell as is often the case for PBE exchange functionals.
However, the VASP-calculated vibrational modes are in qualitatively
  good agreement with measurements from the THz spectroscopy.
We note that our measured low-temperature spectrum itself is in overall good agreement with
previous reports for solid 1,3-DNB.
Our high-resolution measurements enabled us to clearly identify a  low-frequency mode at 0.59 THz and another mode at 0.86 THz.
Quantitatively, the theoretical (VASP)
and experimental spectra are overall
 offset by about 0.4 THz.
While this is of little concern at high frequencies, it makes a direct first-principles prediction of low-frequency modes below 2.5 THz challenging for absolute fingerprinting of molecular crystals. Nevertheless,
the DFPT calculated modes still capture the general trends
in the absorption spectrum of the 1,3-DNB crystal.
In the future, we plan to
 improve the DFT
 calculations by incorporating van der Waals
interactions for the long-range dispersive forces between molecules to correct for the overestimation of the optimized
 unit cell volume and lowest normal modes.

\textit{Acknowledgements.~}
We thank Xiaodong Wen for helpful discussions regarding the DFT calculations and crystallographic databases.
This work was supported by the U.S. DOE under Contract No. DE-AC52-06NA25396 through the LDRD Program.
We gratefully acknowledge computing allocations by NERSC, a U.S. DOE Office of Basic Energy Sciences user facility, under Contract No. DE-AC02-05CH11231.

\bibliography{references}

\end{document}